\documentclass[11 pt, article]{amsart}
\usepackage{amsmath}
\usepackage{latexsym}
\usepackage[all]{xy}
\usepackage{color}

\include{amslatex}

 \numberwithin{equation}{section}

\begin{document}
\begin{title}[On the emergent origin of the inertial mass]
{On the emergent origin of the inertial mass}
\end{title}
\date{\today}
\maketitle
\begin{center}
\author{Ricardo Gallego Torrom\'e}\footnote{Email: rigato39@gmail.com}
\end{center}
\begin{center}
\address{Department of Mathematics\\
Faculty of Mathematics, Natural Sciences and Information Technologies\\
University of Primorska, Koper, Slovenia}
\end{center}
\begin{center}
\author{J. M. Isidro}\footnote{Email: joissan@mat.upv.es}
\end{center}
\begin{center}
\address{Instituto Universitario de Matem\'atica Pura y Aplicada,\\
Universitat Polit$\grave{\textrm{e}}$cnica de Val$\grave{\textrm{e}}$ncia, Valencia 46022, Spain}
\end{center}
\begin{center}
\author{Pedro Fern\'andez de C\'ordoba}\footnote{Email: pfernandez@mat.upv.es}
\end{center}
\begin{center}
\address{Instituto Universitario de Matem\'atica Pura y Aplicada,\\
Universitat Polit$\grave{\textrm{e}}$cnica de Val$\grave{\textrm{e}}$ncia, Valencia 46022, Spain}
\end{center}
\begin{abstract}
 In the context of a particular framework of emergent quantum mechanics, it is argued the emergent origin of the inertial mass of a physical systems. Two main consequences of the theory are discussed: an emergent interpretation of the law of inertia and a derivation of the energy-time uncertainty relation.
\end{abstract}

\section{Introduction}
The problem of finding a consistent picture unifying quantum mechanics and the gravitational interaction, usually in the form of a quantum theory of gravity, has been the object of intense investigations and research in the field of theoretical physics during decades. The most common view is that the gravitational interaction must be quantized. The exact meaning of quantum gravity varies greatly from theory to theory, and one has several candidates for an unifying paradigm, like string theory, loop quantum gravity, non-commutative spacetimes or causal set theory, to mention several remarkable ones. The opposite approach, namely {\it relativize the quantum theory}, has been advocated by R. Penrose \cite{Penrose,Penrose 2005,Penrose 2014a}. Other points of view towards a unifying theory explore the possibility that gravity is emergent, while other fields are of quantum nature. This is the case of Jacobson's theory \cite{JacobsonI}, Verlinde's entropic gravity approach \cite{Verlinde2011,Verlinde2017} and Padmanabhan's approach to classical gravity as a thermodynamical system \cite{Padmanabhan2012a,Padmanabhan2012b,Padmanabhan2015}, three significant proposals for emergent gravity. Remarkably, such emergent approaches to gravity rely on quantum mechanical principles to the extent that a version of Unruh's formula relating temperature and acceleration and that the weak equivalence principle must hold.

Apparently pursuing different questions, emergent approaches to quantum mechanics have been the object of investigation by several authors from the point of view of an underlying fundamental theory \cite{Adler,Hooft2016,Hooft2021,SharmaSingh,DeSinghVarma2019,Singh,Ricardo2014,Elze2003,Elze2009,Dolce 2013}. Other approaches attack the emergent character of quantum theory from a thermodynamical point of view \cite{Dacosta et al.2013,Dacosta et al.2012,FernandezIsidroPerea2013,FernandezIsidroVazquez2016}.

   We discuss the emergent origin of inertial mass in the context of one these emergent approaches to the foundations of quantum mechanics \cite{Ricardo2014}. The fundamental idea of this approach to emergent quantum mechanics is that quantum systems are emergent from an underlying level of physical reality. At such deeper level of description, physical systems are described by deterministic dynamical systems with many degrees of freedom, the dynamics being almost cyclic for quantum systems that do not interact quantum mechanically. In this context, it is shown how the notion of inertial mass has an emergent character, directly linked with the sub-quantum degrees of freedom of the associated Hamilton-Randers system. In particular, a relation between mass and the semi-period of the fundamental cycle is presented. Furthermore, we show how a simple model for the semi-period implies a direct relation between the mass and the number of degrees of freedom $N$. Several consequences of the theory are explored. Among them, a new interpretation of the law of inertia of classical dynamics, linked with the notion of complexity and emergence, and a derivation of an energy-time uncertainty relation, that can be interpreted as a sharper relation than the standard energy-time uncertainty relation of quantum mechanics.

\section{Notion of Hamilton-Randers dynamical systems} The fundamental notions of Hamilton-Randers theory can be found in \cite{Ricardo2014}. Here we distill the necessary notions of the theory to developed the ideas of this paper. One starts by considering a smooth manifold $\widetilde{M}$ as the configuration space of the system.
  The state of a Hamilton-Randers system is described by a point of the co-tangent space $T^*\widetilde{M}$.
In particular, it is assumed a product structure for the configuration manifold of the form
 \begin{align}
\widetilde{M}\cong \,\prod^N_{k=1}\,  T\,M^k_4,
\label{manifoldM}
\end{align}
where all the manifolds $M^k_4,\,k=1,..,N$ are diffeomorphic to each other.
This construction is consistent with the assumption that all sub-quantum degrees of freedom are identical.

In Hamilton-Randers theory each quantum system can be described in terms of an associated Hamilton-Randers dynamical system. Indeed,
Hamilton-Randers theory is based on generic properties that geometric models of Randers type metrics on tangent bundles $T^*TM_k$. A Randers space is a metric space where a Riemannian or pseudo-Riemannian norm function $\alpha$ has been perturbed by a linear term $\beta$.
The Randers metrics provide three fundamental elements: first, a notion of proper time and the corresponding geodesic flow. Second, a notion of measure, that is used below in taking the averaging operation. These two structures play relevant roles in the construction of the theory. Finally, it determines the exact form of the dynamics for the sub-quantum degrees of freedom, in the form of first order ordinary differential equations, a dynamics that was denoted by $U_t$. This dynamical law, which is applied to the fundamental $k$-degrees of freedom, makes use of time parameters, that were called $t$-time. Such parameters are fundamentally different from the time parameters that could appear in the description of quantum mechanics of classical systems.
Indeed, the {\it emergent degrees of freedom} are described by a dynamics denoted $U_\tau$. The emergent degrees of freedom can be seen as a coarse grained average description of the system of $N$ sub-quantum degrees of freedom of a Hamilton-Randers system. Such emergent degrees of freedom are associated with probability densities and it is indeed possible to show that, for systems with a definite number of quantum degrees of freedom, such probability densities are consistent with Born's rule of quantum mechanics \cite{Ricardo2014}. Therefore, the $U_\tau$ dynamics the associated with the quantum dynamics, a result that can be potentially extended to quantum field theories.

The notion of $2$-time dynamics is of fundamental relevance for the construction of the emergence mechanism proposed in Hamilton-Randers theory. The consequences of adopting the $2$-time dynamics and the effective description of the system in terms of the coarse grained description includes a mechanism for the reduction of the wave function \cite{Ricardo2015a} and the interpretation of entanglement and quantum non-locality  \cite{Ricardo2017b}, based upon the formal projection from the description in terms of the $2$-time dynamics to the description in terms of the $1$-time dynamics describing the dynamical evolution of the coarse grained degrees of freedom.
\section{Interpretation of the semi-period $T$ for a particular class of $t$-parameter}
In Hamilton-Randers theory there is no geometric structure defined on the tangent space $TM$ that can be used to determine a natural $t$-time parameter for the $U_t$ dynamics. Indeed, the Hamilton-Randers dynamical models are invariant under positive oriented $t$-time re-parameterizations. However, the freedom in the choice of the $t$-parameter does not preclude the existence of choices of the $t$-parameter that are particularly enlightening. In particular, it can be useful to identify the semi-period $T$ as a characteristic of the system it describes since, given a fixed $t$-time parameter, different systems can have different semi-periods.
In this context, we postulate the existence of $t$-time parameters of the form $[0,2\,T]\subset \,\mathbb{R}$ such that for any Hamilton-Randers system corresponding to a free quantum system, the relation
\begin{align}
 \log\left(\frac{T}{T_{min}}\right)=\,\frac{T_{min}\,m\,c^2}{\hbar}
\label{ETrelation}
\end{align}
holds good.
$T_{min}$ corresponds to the minimal period of the fundamental cycles for any Hamilton-Randers dynamical system. This minimal period must exist, since by assumption, quantum systems contain a finite number of them, then the limiting case when $N=1$ imposes a limit on the complexity of the quantum system. This shows the need for the existence of a $T_{min}$ if the period is directly linked to the number of sub-quantum degrees of freedom. On the grounds of how quantum systems emerge from Hamilton-Randers theory, $N_{min}$ is expected to be much larger than $1$.
The value of $T_{min}$ depends on the choice of the arbitrary $t$-parameter but, once the parameter has been chosen, $T_{min}$ achieves the same universal value for all the Hamilton-Randers systems.

One expects that the parameter $T$ is a measure of the complexity of the system, since the semi-period $T$ increase with the number of components of the physical system.

\section{The parameter {\it m} as a notion of inertial mass}
It is useful to re-cast the relation \eqref{ETrelation} as a definition of the mass parameter $m$ in terms of the semi-period $T$ of the fundamental cycle.
The mass parameter $m$ of a Hamilton-Randers dynamical system with fundamental semi-period  $T$ is postulated to be given by the relation
\begin{align}
m=\,\frac{\hbar}{T_{min}\,c^2}\,\log \left( \frac{T}{T_{min}}\right).
\label{definitionofinertialmass}
\end{align}
{\bf Fundamental properties of the $m$ parameter}. The following are properties of $m$ that are obtained directly from the relation \eqref{definitionofinertialmass}:
\begin{enumerate}
\item \label{atribute} Since the semi-period $T$ is an attribute of the physical system under consideration, the mass parameter $m$ is also an attribute that increases with the {\it complexity of the system}.

\item \label{positivity of m}\label{non negativity of mass m} For any Hamilton-Randers system, the mass parameter $m$ is necessarily non-negative, with the minimum value for the parameter $m$ being equal to $m=\,0$, that corresponds to $T=\,T_{min}$.

\item \label{conservation of m} \label{preservation of mass for a free system} As long as the period $T$ is preserved for each fundamental cycle of the $U_t$ dynamics, the mass parameter $m$ is preserved and remains the same for each cycle of the $U_t$ and the $U_\tau$ dynamics.

\item \label{complexity} Since the value of the semi-period is linked to the ergodic properties of the system, $T$ is a measure of the complexity of the system. Hence $m$ is a measure of the complexity of the system as well.

\end{enumerate}
The properties \ref{atribute}-\ref{conservation of m} make it reasonable to identify the parameter $m$ given by the relation \eqref{definitionofinertialmass} as the mass parameter of the Hamilton-Randers system. From the above points the emergent origin of the mass parameter $m$ given by the relation \eqref{definitionofinertialmass} as a measure of the complexity of the system is also direct.

The above interpretation of the parameter $m$ is associated to a particular class $t$-time parameters characterized by the fact that the semi-period $T$ of the fundamental cycles of a free system is constant. Indeed, given any positive $t$-parameter for the $U_t$ dynamics, one could obviously define $m$ by the relation \eqref{definitionofinertialmass}, but in general, such a parameter $m$ will not be associated with the Hamilton-Randers system, none of the properties \ref{atribute}-\ref{complexity} above will hold good.
\subsection{Inertia as emergent phenomenon}
Within the above context of associating the mass parameter $m$ with the complexity of the system and the emergence of the notion of mass, it is reasonable to expect that coherent changes in the state of the system will be more difficult to be continuously kept as the complexity of the system increases. In the theory developed above, the mass parameter $m$ as given by the relation \eqref{definitionofinertialmass} is related to the complexity of the fundamental cycles, which is related to the semi-period: the greater the complexity of the system, the larger the mass parameter $m$. Hence the parameter $m$ is also becomes a measure of the opposition to coherent changes in the state of the system. Thus the tendency of a classical or quantum system is to {\it resist} change in its state, and the resistance of change is greater when the mass $m$ is larger. Therefore, the emergent nature of {\it The Inertial Law} and the identification of $m$ with the inertial mass becomes apparent.

\section{Periods and mass parameter for composite Hamilton-Randers systems}
Let us consider two arbitrary Hamilton-Randers systems $a,\,b$ with semi-periods $T_a$ and $T_b$. When the sub-systems $a$ and $b$ do not interact between each other, or if they interact, the effect of the interactions can be neglected, the {\it joint Hamilton-Randers system} $a\sqcup b$ describing the joint sub-systems $a$ and $b$ can be defined as follows. From the point of view of the quantum mechanical description, such physical systems are described by elements of the tensor product of the corresponding Hilbert spaces. From the point of view of Hamilton-Randers theory, if the sub-quantum molecules determining the quantum system $a$ do not interact with the sub-quantum molecules defining the quantum system $b$ and the corresponding structures are $(M_a, (\alpha_a,\beta_a))$ and $(M_b,(\alpha_b,\beta_b))$ respectively, then there is a natural Randers structure constructed from the structures $a$ and $b$ which is the product of {\it Randers structure},
   \begin{align*}
   \left(M_a\times M_b,\, \left(\alpha_a\otimes \alpha_b,\,\beta_a\otimes \beta_b \right)\right).
   \end{align*}
This structure defines the Hamilton-Randers system $a\sqcup b$.

If the systems $a$ and $b$ do not interact and the corresponding semi-periods are $T_a$ and $T_b$, then we further can assume that the semi-period for $a\,\sqcup\,b$ is given by the multiplicative rule,
 \begin{align}
T_{a \,\sqcup\, b}:=\,{T_a}\, {T_b}.
 \label{compatibilityconditionabwhennotmeasurable}
 \end{align}
 Note that this rule is not satisfied for a generic $t$-time parameter, since for the joint system $a\sqcup b$ one can choose an arbitrary $t$-time parameter to describe the $U_t$ dynamics. Conversely, the condition \eqref{compatibilityconditionabwhennotmeasurable} can be taken as the definition of non-interacting systems at the quantum mechanical level.

The above discussion shows two general features of the theory:
\begin{itemize}

\item The semi-period associated with a composite system should increase exponentially with the number of independent, non-interacting components of the system.

\item From the relation \eqref{compatibilityconditionabwhennotmeasurable} the mass $M$ and the semi-period $T$ given by the relation \eqref{ETrelation} and such that the $t$-parameters such that $T_{min}=1$, it follows that
the mass parameter of a composite non-interacting system $a\sqcup b$ is additive,
\begin{align}
 M_{a \,\sqcup\, b}=\,M_a+\,M_b.
 \label{additivityofmass}
 \end{align}
\end{itemize}

\subsection{Models for the semi-periods of the internal dynamics} Although the $U_t$ dynamics is manifestly $t$-time re-parametrization invariant (see \cite{Ricardo2014}, {\it Chapter 3}), the interesting features of the $t$-parameters for which the relations \eqref{ETrelation}, \eqref{compatibilityconditionabwhennotmeasurable} \eqref{additivityofmass} hold good invite to consider the construction of specific models realizing such a class of $t$-time parameters.

 A fundamental characteristic of a Hamilton-Randers system is the number of sub-quantum degrees of freedom $N$. Also, we would like to point out that the dynamics $U_t$ preserves $N$. Thus $N(t)$ is constant for any choice of the $t$-parameter. In a process where the system can be subdivided into two parts, the number $N$ is also portioned: if there is a process of the form $1\to 2\sqcup 3$, then we assume the condition
\begin{align}
 N_1=\,N_2+N_3\,-N_{min}.
 \label{preservation of the number of lines}
\end{align}
Thus the number of degrees of freedom of a composite system is the sum of the independent degrees of freedom; then $N_{min}$ appears as the number of degrees of freedom of a common border and is subtracted in order to do not count twice such degrees of freedom.

In the following, we consider two models relating the semi-period $T$ as function of the number of degrees of freedom $N$.
\bigskip
\\
{\bf Model 1}. Let us consider the following model for the semi-period,
\begin{align}
\frac{T}{T_{min}}:=\,\frac{T(N)}{T_{min}} =\,N^{\epsilon},
\label{model 1 for semiperiod}
\end{align}
where $\epsilon$ is a positive constant that does not depend upon the specific Hamilton-Randers  system, nor on  the number of sub-quantum degrees of freedom $N_a$. Then it is clear that $T$ is a measure of the complexity of the system. The model \eqref{model 1 for semiperiod} for the semi-period is consistent with the relation \eqref{ETrelation}, but does not reproduce the multiplication rule \eqref{compatibilityconditionabwhennotmeasurable} when the relation $N_1=\,N_2+N_3-\,N_{min}$ holds good.
\bigskip
\\
{\bf Model 2}.  A model that it is consistent with \eqref{preservation of the number of lines} is the exponential model
\begin{align}
T=\,T_{min}\,\exp{\left(\frac{\lambda\, T_{min}\,c^2}{\hbar}\left(N-N_{min}\right)\right) },
\label{model 2 for semiperiod}
\end{align}
where $\lambda$ is a constant relating the mass parameter $m$ and the number of degrees of freedom:
\begin{align}
m = \, \lambda\left(N-\,N_{min}\right).
\label{relacion masa con N}
\end{align}
In the model given by  \eqref{model 2 for semiperiod} $N_{min}$ corresponds to $T_{min}$.
Furthermore, when $T= \,T_{min}$, then $m=0$ and in addition if $T_{min}=\,1$, then the model \eqref{model 2 for semiperiod} is such that $T_{a\sqcup b}=T_a\,\cdot T_b$.
According to this model, the mass parameter $m$ is equivalent to the number $N$ of sub-quantum degrees of freedom, making therefore $m$ a measure of the {\it quantity of sub-quantum matter}, in analogy of the notion of mass as amount of matter in classical mechanics.
Furthermore, from the definition \eqref{definitionofinertialmass} of the mass parameter $m$ and the relation \eqref{relacion masa con N}, preservation of the number of lines $N$ in the sense of the relation \eqref{preservation of the number of lines} implies additivity of the mass parameter $m$. These properties makes Model 2 a very appropriate model for the semi-period $T$.

\section{Emergence of an energy-time uncertainty relation from emergent quantum mechanics}
The relation \eqref{ETrelation} is not equivalent to the quantum energy-time uncertainty relation, since the $t$-parameter is not an external time parameter. Also, let us note that the relation between $m$ and $T$ is given by $\log T$ rather than being linear with $T^{-1}$, as should be expected for an energy-time uncertainty quantum relation. Furthermore, according to the relation \eqref{ETrelation}, $m$ increases monotonically with $T$.

  However, if we consider the variation of the parameter $m$ due to a variation of the period $2\,T$ in the relation \eqref{ETrelation}, we have that
 \begin{align*}
 \Delta \left(m\,c^2\right)=\,(\Delta m)\,c^2 =\,\frac{\hbar}{T_{min}}\,\frac{\Delta{T}}{T}.
 \end{align*}
The variation in the mass can be conceived as a variation due to a continuous interaction of the system with the environment; $\Delta T/T_{min}$ is the number of fundamental dynamical cycles that contribute to the stability of the quantum system.
 In a theory with maximal acceleration spacetime geometry, as we assume to be the case of the models $(M^k_4, \eta,\beta)$ for each sub-quantum degree of freedom under consideration, the expression  $m\,c^2$ is the energy of a system measured by an observer instantaneously co-moving with the system \cite{Gallego-Torrome2019}, when the system has zero proper acceleration.  If there is a local coordinate system associated with the Hamilton-Randers system in a way which is in some sense instantaneously co-moving with the system, then we can apply the relativistic expression for the energy \cite{Gallego-Torrome2019}. The value of the semi-period $T$ is a characteristic of the quantum system associated in such a local coordinate frame.
 From the definition of $T_{min}$ we have that $\Delta{T}/T_{min}\geq 1$. Then
 \begin{align}
 \Delta E\,T\geq \hbar
 \label{Energytimeuncertainty}
 \end{align}
 with the {\it spread of energy} defined as
 \begin{align}
 \Delta E :=\,\Delta(m\,c^2)=\,\Delta(m)\,c^2.
 \label{spread of energy}
 \end{align}
 Therefore, the {\it uncertainty} in the energy at rest $E$ associated  with the system is related to the inverse $T^{-1}$ of the semi-period.

One natural interpretation of the energy-time relation \eqref{Energytimeuncertainty} is to think the quantity $\Delta E$  as the minimal exchange of energy between the Hamilton-Randers system and the environment in such a way that the system is stable at least during a whole cycle of semi-period $T$.
This energy exchange is measured in an instantaneous inertial reference frame in co-motion with the system just before the system changes to another different state or it decays to another different type of quantum system.
Identifying the $t$-time parameter describing the $U_t$ evolution of the sub-quantum degrees of freedom with a macroscopic coordinate time of a co-moving system with the quantum system and assuming that the average lifetime $\tau$ is much longer than $T$, one obtains the standard energy-time quantum uncertainty relation,
   \begin{align}
    \Delta E\,\tau\geq \hbar .
    \label{Energytimeuncertainty 2}
   \end{align}
This interpretation is fully consistent with $T$ as a parameter associated with the $U_t$ evolution and at the same time, an intrinsic parameter associated to the system.

\section{Discussion}
In this paper the relation between certain $t$-time parameters and the emergent origin of mass has been discussed in the context of Hamilton-Randers theory \cite{Ricardo2014}. According to such framework, it is essential to consider the $2$-time character of the dynamics. Dynamical systems where time is represented by two parameters exist in the literature. In classical dynamics, fast/slow variables are defined, just to split in a convenient way the dynamics of different classes of degrees of freedom. In such models, there is a diffeomorphism between the different time parameters \cite{Arnold}, so indeed, there is only one time flow, but labelled by different parameters. More related to our theme, this type of fast/slow dynamics appears in the theory of emergent quantum mechanics according to 't Hooft \cite{Hooft2016,Hooft2021}. However, such a diffeomorphic relation between the $t$-time parameters and the $\tau$-time parameters is absent in Hamilton-Randers theory, because the time parameters used in the quantum dynamics have an emergent character \cite{Ricardo2014,Ricardo general dynamics} and are essentially discrete (although approximated by a continuum), while the $t$-parameters do not have such an emergent origin. Let us also mention that the notion of $2$-dimensional time that we use is different from the $2$-dimensional time that occurs in $2T$-time physics models of I. Bars \cite{Bars2001, Bars2007,Bars2008}. In Bars' theory, the $2$-dimensional character of time is embedded in the signature of the fundamental $4+2$ dimensional pseudo-Riemannian arena, while in Hamilton-Randers theory such a link with the geometry of the spacetime is in principle absent. This is because of the emergent character of $\tau$-time parameters from the cyclic structure of the underlaying $U_t$ dynamics.

The relation between the expression \eqref{preservation of the number of lines} and the relation \eqref{definitionofinertialmass} leads to an interesting interpretation of massless systems. From one side, the relation \eqref{preservation of the number of lines} can be interpreted as associating $N_{min}$ to the number of degrees of freedom in {\it borders}: when a system $S_1$ is joined with system $S_2$, the degrees of freedom at the topological border of the joint region should be {\it counted} only once. Hence one needs to subtract the degrees of freedom at the joint region, namely $N_{min}$. On the other side, the number of degrees of freedom corresponds to a massless quantum system, according to the relation \eqref{definitionofinertialmass} and the model of the semi-period \eqref{model 2 for semiperiod}. This implies that massless quantum particles should be interpreted as {\it borders} between {\it bulks} in $T^*TM$ and that for all massless systems have associated the same number of sub-quantum degrees of freedom, namely $N_{min}$.

It is also very suggestive to compare the relation \eqref{definitionofinertialmass},
\begin{align*}
m=\,\frac{\hbar}{T_{min}\,c^2}\,\log \left( \frac{T}{T_{min}}\right),
\end{align*}
with the expression of the entropy in terms of microstates,
\begin{align*}
S=\,k_B\,\log V,
\end{align*}
where $k_B$ is the Boltzmann constant and $V$ is the phase space volume compatible with the {\it coarse grained} description of the system. To relate $m$ and $S$ given by the above expressions implies to relate the quantities $T$ with $V$. Let us note that while the mass of a quantum system is to be expected to be preserved, if the system evolves freely, the entropy is only preserved when it is maximal. Therefore, an entropic interpretation of mass based on the relation of $T$ with $V$ is subjected to this constraint.
On the other hand, when thinking in terms of sub-quantum degrees of freedom, a dimensional argument suggests that
\begin{align}
T\sim\,\c\, \tilde{V}^{1/16},
\label{relacion m to V}
\end{align}
where $\tilde{V}$ is the volume of the phase space of each sub-quantum degree of freedom.
A maximum entropy state $\tilde{S}$ corresponds to stable systems, for any Hamilton-Randers system. If \eqref{relacion m to V} holds good and because $\tilde{V}$ is a geometric invariant, then there is an emergent and general covariant definition of mass of the system,
 \begin{align}
 \tilde{m}=\,\frac{1}{16}\,k_B\,N\,\log \frac{\tilde{V}}{\tilde{V}_{min}},
 \end{align}
where $V_{min}$ corresponds to $T_{min}$.
 One is then tempted to re-define the entropy of the system in the form
 \begin{align*}
 \tilde{S}=\,\,k_B\,\,N\,\log \tilde{V},
 \end{align*}
providing an entropic character to the mass $\tilde{m}$. Note that $\widetilde{S}$ corresponds to the entropy of $N$ independent sub-quantum degrees of freedom. However, the condition of independent (or non-interacting) sub-quantum degrees of freedom is indeed not adequate, due to the highly non-trivial $U_t$ dynamics.

 Finally, we should remark that, given the emergent nature of inertial mass and since the weak equivalence principle holds good in emergent quantum mechanics \cite{Ricardo2014,Ricardo 2019a}, then the passive gravitational mass must be also emergent, providing another argument in favour of the emergent character of gravity, consistently with the general mechanism found in the context of Hamilton-Randers theory \cite{Ricardo2014,Ricardo 2019a}.

\section{Conclusion}
An emergent concept of inertial mass has been discussed in the framework of Hamilton-Randers systems. The idea brings to light a new perspective on several fundamental notions, among them the law of inertia and the energy-time uncertainty relation. Furthermore, the emergent nature of inertial mass, together with the inertial mass/gravitational mass equality, implies an additional argument in favour of the emergent character of gravity. Indeed, the ideas described in this paper are part of a more ample program of research where fundamental notions of current physics are viewed as emergent notions from a fundamental theory beneath quantum theory \cite{Ricardo2014}.

\end{document}